The Stern-Gerlach Experiment-*Was Sind und was Sollen*
Jeremy Bernstein
Stevens Institute of Technology

In 1888 the German mathematician Richard Dedekind published a paper which he called *Was Sind und Was Sollen die Zahlen* in which he described the deep significance of ordinary numbers-*Zahlen*. The purpose of this essay is to explain the deep significance of Stern-Gerlach experiments-real and imagined. Feynman once noted that all of quantum mechanics is summarized in the double slit experiments. I will try to persuade you that Feynman was wrong-all of quantum mechanics is summarized in Stern-Gerlach experiments-at least all of quantum mechanics that is really mysterious. I begin by reminding you of the actual Stern-Gerlach experiment beginning with a biographical note.

I. History

Otto Stern was born in 1888 in Sobrau in Prussia. He took his doctorate in physical chemistry at the University of Breslau in 1912 but then he went to Prague so that he could study with Einstein who was for a brief time at the German University. It is reported that their discussions took place in a café attached to a brothel.[1] Einstein persuaded Stern to switch his interests to physics and, indeed Stern followed Einstein to Zurich when Einstein took a position at the ETH. The First World War intervened and Stern served in the German army. He was nonetheless able to do some physics including the study of the separation of isotopes using gaseous diffusion. This was one of the methods used at Oak Ridge to separate the uranium isotopes that went into the Hiroshima bomb. After the war Stern became Max Born's assistant in Frankfurt. Born was interested in molecular beams, something that Stern was already interested in ,and he encouraged Stern to continue in this direction.

---

[1] For a very nice discussion of the history of the Stern-Gerlach experiment, including this tidbit see "Stern and Gerlach: How a Bad Cigar Helped Reorient Atomic Physics" by Bretislav Friedrich and Dudley Herschbach, Physics Today, December 2003,p.53-59.. For a detailed account see The Historical Development of Quantum Theory, Volume 1,Part 2 by Jagdish Mehra and Helmut Rechenberg. Springer-Verlag, New York, 2000.

Walther Gerlach was born in 1889 in Biebrich. He received his education which focused on experimental physics in Tübingen where he took his doctorate in 1912. He too served in the war and afterwards had various positions including one in industry. But in 1920 he also went to Frankfurt where he obtained a position at the Institute for Experimental Physics which was in the same building as Born's group . Gerlach had had interests in molecular beam experiments and so it was natural that he and Stern should join forces. It was Stern who hit on the idea for the experiment that would immortalize them both. Stern realized that if you could make an inhomogeneous magnetic field of sufficient strength it would act on the magnetic dipole moment of an atom and cause it to change its trajectory. He had computed that he would need about $10^4$ Oersted per centimeter to do the job and he asked Gerlach whether he could make a magnetic field like this. Gerlach said this would not be difficult and, indeed, he could do even better. Stern had figured that if this field gradient could be held constant for about 3 centimeters then a deflection of something like $10^{-2}$ millimeters would be produced providing all on this took place in a very good vacuum.

Stern was absolutely clear on what he wanted this experiment to demonstrate and indeed in 1921 he published a paper "Ein Weg zur experimentellen Pruefung der Richtungsquantelung im Magnetfeld"[2] which said it all: "A Method Using a Magnetic Field to Demonstrate Space Quantization." I have taken a little liberty with the translation of "*richtungs*" but "space quantization" is how it was universally referred to. By 1921, when Stern wrote this paper, it was generally accepted that the atom consisted of a tiny nucleus surrounded by a planetary cloud of electrons. Most physicists would probably have agreed that these electrons moved in selected Bohr orbits. No one had seen an electron moving in such an orbit so the whole idea seemed rather abstract. But if it was true it had important consequences for the intrinsic angular momenta of such atoms. Ignoring any contributions from the nuclei this angular momentum would be produced by the orbital angular momenta of the circulating electrons. Spin was unknown. In the "old quantum theory", because angular momentum was quantized, the planes of these orbits could tilt at only discrete angles with respect to the direction of the magnetic field. This is what was known as "space quantization." In the case of the silver atom, for example, there could in principle be only three directions which we can designate as ±1 and 0. The zero direction was one in which the plane of the orbit and the direction of the magnetic field were parallel. But Bohr had argued that this direction should be excluded since the orbit would be unstable. Stern accepted this argument which implied that the angular momentum could point in only two directions. However, the magnetic force is proportional to the magnetic moment of the atom and this is, in this picture, proportional to the orbital angular momentum of the electrons. Thus this force, Bohr claimed, can point in only two directions which meant that the magnetic field could split the incident atoms into two and only two beams. This was the prediction that Stern proposed to test.

---

[2] Stern, O. (1921). "Ein Weg zur experimentellen Pruefung der Richtungsquantelung im Magnetfeld". *Zeitschrift für Physik* **7**: 249-253..

    The experiment turned out to be both very difficult to perform and to finance. Born tried to help out by offering a paid lecture on Einstein's theory of relativity and turning the proceeds over to the experimenters. He also mentioned the problem to a friend who was going to visit the United States. The friend solicited Henry Goldman, whose father Marcus, an immigrant from Frankfurt, had founded the firm that became Goldman Sachs. A check for a few hundred dollars was forthcoming which saved their experiment. One wonders what mister Goldman had grasped about space quantization. In 1921, Stern left Frankfurt for Rostock and Gerlach remained working on the experiment alone. Prior to that the two of them had had an incident with cigars. Their detectors were plates on which the silver was deposited. At first they saw nothing but the cheap cigars they were chain smoking had a high sulfur content. The smoke from the cigars deposited sulfur on the plates and the silver beam showed up,

After that they used photographic plates. By 1921, Gerlach had seen the deposit from the two beams as predicted. They then did careful repeat experiments in Frankfurt which enabled them to measure the magnetic moment of the silver atom to something like a ten percent accuracy. In 1922 they published a joint paper. "*Das magnetische Moment des Silberatoms*". It is intersting to note that In this paper the authors list themselves alphabetically- Gerlach and Stern.

Pasteur once famously remarked that in scientific research chance favors the prepared mind. Perhaps in the case of Stern and Gerlach one might cite the cigars. I am inclined to put this in the *se non è vero, è ben trovato* category. In any event they switched to using photographic plates. A much better case can be made for their choice of element-silver. If, for example, they had been able to use yttrium they would have found four lines which would have been incomprehensible.[3] We now understand, which Stern and Gerlach could not have, why they got their result. It took until 1925 before Samuel Goudsmit and George Uhlenbeck proposed electron spin. This was necessary but not sufficient. Bohr's model for silver had all the electrons in saturated shells which had in total no angular momentum. Outside these shells was a single valence electron. But it took until 1927 until the Scottish physicist Ronald G.J.Fraser argued in a brief afterthought to a paper[4] that to fit the data on the magnetic moments of some atoms one needed to assume that the valence electron was in an S-state-no orbital angular momentum-so that the total angular momentum of the atom was coming from the spin of the electron and hence the magnetic moments should be approximately that of the electron. In short, unknown to them, Stern and Gerlach had measured the spin of the electron.

Before I turn to the quantum mechanics of the Stern-Gerlach experiment I want to fill out briefly the arc of their lives. Stern moved to the

---

[3] The angular momentum iof the atom s 3/..
[4] Proc. R.Soc. A **114**,212, (1927)

University of Hamburg in 1923 where he remained until 1933 when the Nazi racial laws forced him to leave Germany. He took a position at the Carnegie Institute of Technology in Pittsburgh where he remained until his retirement in 1945. He then became a professor emeritus at the University of California at Berkeley. He died in Berkeley in 1969. In 1943, he was awarded the Nobel Prize in Physics. The citation is interesting. It says that the prize was "for his contribution of the molecular ray method and his discovery of the magnetic moment of the proton." Of the Stern-Gerlach experiment there is no mention.

Gerlach's arc is very different. In 1925 he became a professor at the University of Munich. He held this position all during the Nazi regime. It seems that Gerlach was never a member of the Nazi Party but apparently he had a brother who was in the SS. After fission was discovered, in conjunction with the German army, there was a program to exploit nuclear energy which Gerlach took part in along with people like Heisenberg. In 1944 became the director of the program. But in May of 1945 he was captured by the mission to which Goudsmit was the scientific advisor. Gerlach was flown to England and spent six months confined in a manor house near Cambridge, Farm Hall, with nine other nuclear scientists including Heisenberg. Their conversations were recorded and make fascinating reading as does the evaluation the British made of their detainees. Of Gerlach they commented, "Has always been very cheerful, but from his monitored conversations is open to suspicion because of his connections with the Gestapo. As the man appointed by the German government to organize the research on uranium, he considers himself in the position of a defeated general and appeared to be contemplating suicide when the announcement [of the bombing of Hiroshima] was made."[5] In 1948, Gerlach became a professor again in Munich where he remained until 1957 after which he had various jobs in the German scientific establishment. He died in 1979 in Munich. I think it is fair to say that his work with Stern was the highpoint of his scientific career.

---

[5] Hitler's Uranium Club, annotated by Jeremy Bernstein, Copernicus Books, New York, 2001, p.143.

II: Quantum Theory

In preparing this essay I looked at several well-known quantum mechanics texts to see how they treated the Stern-Gerlach experiment. As it turned out this was an easy task because in nearly all the texts I looked at, it was either not mentioned at all, or got a page or two. A notable exception was Kurt Gottfried's *Quantum Mechanics*[6] where the Stern-Gerlach experiment is treated in some detail and some of the lessons drawn. Another notable exception is *Quantum Theory* by David Bohm.[7] This is the treatment I am going to follow if for no other reason than I used this text when I was a student. Also what Bohm did in this treatment has influenced physicists as diverse as John Bell and Julian Schwinger. Bohm's book has an odd history. As he writes in the preface, it was influenced by lectures that Robert Oppenheimer gave at Berkeley about quantum theory. Bohm later reported that when he went to see Oppenheimer about publishing the book, instead of encouraging him Oppenheimer said that he shouldn't without giving any reason-another of Oppenheimer's quixotic reactions. But more than this, by the time the book came out, Bohm had serious misgivings about the very interpretation of the theory that he was brilliantly defending in the book. Indeed, he produced a "pilot wave" interpretation, something that had first been done in the late 1920s by Louis de Broglie who had abandoned it. In Bohm's text there is not the slightest hint of any of this.

We shall with Bohm assume that the Stern-Gerlach magnetic field is oriented in the z direction. The magnet will influence our spinning particle to move in that direction-or its opposite-and from this motion we can determine the spin. Before any interaction with the magnet the wave function $\psi_o$ will be of the form, using Bohm's notation[8]

$\psi_o = f(z)_0 (c_+ v_+ + c_- v_-)$.

---

[6] A 1989 edition was published by Westview Press, Boulder, Colorado
[7] Prentice Hall, New York, 1951.
[8] This treatment can be found in Bohm op.cit. Chapter 22.

Here $v_\pm$ are the spin functions associated with the two directions of spin and $c_\pm$ are complex numbers and $f(z)_o$ is the spatial part of the wave function. In short we are assuming that the wave function is initially a coherent superposition of the two possible spin states. As long as the wave function evolves with a Schrödinger equation, it will remain a coherent superposition but with different coefficients. At a later time in the presence of the interaction with the magnetic field the wave function becomes

$$\Psi(z,t)=f(z,t)_+ v_+ + f(z,t)_- v_-.$$

Ever since the work of von Neumann[9] it has been customary to describe a measurement of a quantum mechanical system in interaction with an apparatus such as a Stern-Gerlach magnet by using an impulse approximation. This assumes that the interaction is intense and of short duration. In the Stern-Gerlach experiment the silver was heated in a furnace to about 1300K. The atoms emerge through slits which collimates them. Stern and Gerlach verified that their velocity distribution was Maxwellian. Typical speeds were about one kilometer a second. The length of their magnetic field was about 10 centimeters hence an atom spent about $10^{-4}$ seconds in it-an impulsive interaction for certain.

The interaction part of the Hamiltonian, $H_I$, during this period dominates the Hamiltonian. Thus the Schrödinger equation is approximately given by

$$i\hbar \partial \psi/\partial t = H_I \psi.$$

The interaction is given by[10]

$$H_I = \mu \boldsymbol{\sigma} \cdot \mathbf{H}.$$

---

[9] For the English translation see Mathematical Foundations of Quantum Mechanics, Princeton University Press, Princeton, 1955 Chapter V.
[10] Following Bohm I use **H** rather than **B**.

Here $\mu=-e\hbar/2mc$, the Bohr magneton for the negatively charged electron and **σ** are the Pauli spin matrices while **H** is the magnetic field vector. In the Stern-Gerlach experiment the field points in the z direction. One expands this component taking into account only the first degree in the inhomogeneity; ie,

$$H_z \approx H_0 + H_0' z$$

where

$$H_0' = (\partial H_z/\partial z)_{z=0}.$$

Using the expansion of the wave function given above we have a pair of Schrödinger equations[11]

$$i\hbar \partial f(z,t)_+/\partial t = \mu(H_0 + H_0' z) f(z,t)_+$$

and

$$i\hbar \partial f(z,t)_-/\partial t = -\mu(H_0 + H_0' z) f(z,t)_-.$$

The difference in sign represents the orientation of the z component of the spin. The boundary conditions are at t=0

$$f_\pm = f_0(z) c_\pm,$$

so the solutions are

$$f_+ = c_+ f_0(z) \exp(-i\mu(H_0 + H_0' z)t/\hbar)$$

and

---

[11] It takes some care to satisfy the conditions that both $\nabla \times H = 0$ and $\Delta \cdot H = 0$. For a discussion of this see for example D.E.Platt. Am.J.Phys. **60**,306-308, (1992). The latter condition implies that there are deflections in other than the z direction which I am, as is traditional in these discussions, ignoring.

$$f_-= c_- f_0(z)\exp(i\mu(H_0+H_0'z)t/\hbar)$$

so

$$\psi = f_0(z)(c_+\exp{-(i\mu(H_0+H_0'z)t/\hbar)}v_+ + c_-\exp(i\mu(H_0+H_0'z)t/\hbar)v_-)$$

Bohm, using a semi-classical argument,[12] shows that after the particles have traversed the magnet for which it takes a time $\Delta t = L/v$, where L is the length of the magnet and v is the particle speed, two wave packets are formed obeying the classical equations of motion

$$z = \pm H_0'\mu\Delta t t/\hbar.$$

This picture conforms to diagram one sees in the text books that do devote some space to the Stern-Gerlach experiment. These diagrams invariably show two rays emanating at an angle from the magnets. But we must be careful. This is quantum mechanics after all.

To give an idea of why caution is needed here is a brief account of a "paradox" that was invented by Neville Mott in 1929. Mott considered an alpha-particle emitter in the middle of a cloud chamber. Cloud chambers where then the particle detectors *du jour*. Mott supposed that these alpha-particles were emitted in S-waves. S-wave, wave functions have the property that if one takes the surface of any sphere drawn around the emitter, then the wave function has the same value everywhere on that surface. This might naively suggest that an emitted alpha-particle produces ionization droplets everywhere in the cloud chamber. This is not at all what happens. Each alpha-particle produces a straight line track of droplets. What the wave function tells us, and this is all it tells us, is that the angular distribution of these tracks is isotropic-equi-probable in every direction. The straight line comes from an analysis of the interaction of the alpha-particle with its cloud chamber environment. This tells us that we must be careful in interpreting the meaning of a wave function.

---

[12] Bohm, op.cit, p .596 et seq

As I will now explain, in the Stern-Gerlach situation what matters is the "decoherence" of the wave function and this may, or may not, be related to the spatial separation of the wave packets. Why does this matter? Eventually we want to assign a probability to the occurrence of one wave packet or the other. But so long as the wave function is a coherent superposition the two parts of the wave function can interfere with each other so that probabilities of individual wave packets cannot be defined. One can re-phrase this in terms of the density matrix. In this case this is a 2x2 matrix whose components are found by multiplying together the coefficients of the various spin up and spin down combinations. Prior to the interaction with the magnet the matrix will look like $\begin{pmatrix} |c_+|^2 & c_+ c_-^* \\ c_- c_+^* & |c_-|^2 \end{pmatrix}$. Note the off diagonal terms. During the interaction with the magnet these off diagonal terms are replaced by new ones reflecting the interaction. But the probability interpretation demands that this matrix is diagonal so that the diagonal elements are the probabilities. In most contexts in which decoherence is discussed, it is the interaction of the system with its environment which renders the off-diagonal elements small. The environment might consist of the photons left over from the Big Bang. These can decohere a macroscopic system so quickly that we are never aware of these quantum effects. But here all we have is the Stern-Gerlach magnet. If there is going to be decoherence the interaction with the magnet had better bring it about.

Bohm considers this matter although he does not employ the term "decoherence." He notes that if these interference terms are not negligible the whole notion of measurement become impossible. He then argues that in this case the interference terms are negligible. His discussion will look very familiar to anyone who has followed the recent literature on this subject. The interference terms involve products of exponentials and if the arguments of these exponentials are very large then they oscillate so rapidly that effectively they average out to zero. I would not like to try to justify the mathematical rigor of such an argument but it gives the "right" answer. During the encounter with the magnet the argument is approximately $\mu(H_0'z)t/\hbar$. I have dropped the $H_0$ term since this would only make the argument larger. We want to evaluate this after the molecule has gone through the magnet. Bohm takes a magnet of 10cm and a speed of the molecules of $10^4$ cm/sec so the time it takes to go through the magnet is about $10^{-3}$ seconds. He takes $H_0'$ to be 10,000 gauss per cm. The quantity e/mc is about $10^7$/gauss. Thus the argument of the exponential is huge and the decoherence is assured to a high degree of approximation.

Bohm raises an interesting question of principle. Since the decoherence is not complete could one at this stage put in a second Stern-Gerlach magnet and reconstruct the initial coherent state? He notes that if the second magnet was absolutely identical to the first one could probably do it. Schwinger and his collaborators have studied this question In detail.[13] They conclude that to reconstruct the coherent wave function with a 99% accuracy you would have to control the magnets to a one part in $10^5$ accuracy and that a 100% reconstruction is impossible. Like Humpty-Dumpty, you cannot put back the pieces.

It is very important to understand that while decoherence is necessary to the measurement process it is not sufficient.[14] The actual recorded measurement projects out a component of the coherent wave function. The two components are orthogonal to each other which implies that they cannot have evolved in time via a single unitary transformation such as exp(i/ℏHt). If they did then

---

[13] See for example "Is Spin Coherence Like Humpty Dumpty? I. Simplified Treatment" by Berthold-Georg Englert, Julian Schwinger and Marian O.Scully, Foundations of Physics, Vol.18,No.10, 1988. This paper also considers with some care the role of the magnetic forces in other than the z-direction.

[14] See for example, "Why Decoherence has not Solved the Measurement Problem: A Response to P.W.Anderson", by Stephen L.Adler, arXiv:quant-ph/0112095v3 10 May 2002

$|A\rangle = U|O\rangle$ and $|B\rangle = U|O\rangle$ so that $\langle A|B\rangle = \langle 0|U^\dagger U|0\rangle = 1$, a contradiction. This is often referred to as the "measurement problem" since measurements cannot be described by a reversible Schrödinger equation evolution in time. Something new has to be added. This is not the place to try to describe the various attempts to resolve this except to note that once such a measurement has been recorded, the past, at least in the usual interpretation of the quantum theory, cannot be recovered.

In 1935, Einstein, Boris Podolsky and Nathan Rosen published a paper entitled "Can Quantum-Mechanical Description of Physical Reality be Considered Complete?"[15] They imagined two particles that have interacted in such a way that momentum and relative position are conserved. Because of the interaction the wave function of the two particles cannot be written as the product of wave functions for the individual particles. This is something that Schrödinger gave the name "entanglement." They then imagined measuring the momentum of one of the particles. The entanglement then tells them that if they had measured the momentum of the other distant particle they could predict the answer with certainty. They argue that in fact they "know" this momentum even without making the measurement. There is, in their language, an element of "reality" associated with this momentum. Likewise in their set up the position of the second particle is implicitly determined and an element of reality is assigned to it. But Heisenberg's uncertainty principle prohibits a knowledge of the momentum and hence quantum mechanics cannot describe this "reality." The theory is not "complete." Bohr was quick to point out that the only measurements that count in the quantum theory are actual measurements made with an apparatus and not virtual measurements of the kind that were described in the EPR paper. More radically, the second particle does not **have** a momentum or a position prior to such a measurement. There is no paradox if one follows the rules.

---

[15] Phys. Rev.**42**, 777 (1935) I am very grateful to Arthur Fine for critical remarks on the EPR.

It seems as if the actual writing of the paper was done by Podolsky. Einstein later complained that it was too complicated in their paper. Bohm must have thought so too because he invented a version with the Stern-Gerlach set up. It is in his book and it has been with us ever since. Einstein told Bohm after reading his book that it contained the best arguments against his interpretation of the theory that he had seen. Ironically, the arguments did not convince Bohm because, as I have noted, he began the construction of his pilot wave interpretation. In Bohm's version he imagines a source that produces two spin-1/2 particles in a singlet state. The normalized spin part of this wave function we can represent by ($\uparrow_1\downarrow_2 - \downarrow_1\uparrow_2$)/√2. The arrows represent whether the z-component of the spin is pointing up or down. When I discuss Bell shortly I will describe in more detail the correlations entailed when the spin direction of one particle is compared to that of the other. At this point I will note that, with this wave function, each observer at his or her magnet will observe a random pattern of "ups" and "downs", but when these patterns are later compared there will be a perfect anti-correlation of "ups" and "downs", provided that the two magnets produce fields that point in the same direction. Each "up" at one magnet will be associated with a "down" at the other. This is true no matter how far away one magnet is from the other something that Einstein referred to as "spooky actions at a distance."

Here is how the EPR experiment is formulated in this context. We first measure the spin of one of the particles with one of the magnets. If the spin is up the wave function will collapse to $\uparrow_1\downarrow_2$. Following EPR we would argue that we do not have to measure the z-component of the spin of the other particle because we "know" that it is down. So instead we will measure the x-component. We will do this by rotating the magnet through an angle ϴ about the y-axis. The spinor rotation matrix for this case is given by

$\begin{pmatrix} \cos(\theta/2) & \sin(\theta/2) \\ -\sin(\theta/2) & \cos(\theta/2) \end{pmatrix}$. If we apply this to the spin down vector $\begin{pmatrix} 0 \\ 1 \end{pmatrix}$ we arrive at the vector $\begin{pmatrix} \sin(\theta/2) \\ \cos(\theta/2) \end{pmatrix}$. We will let the x component of the Pauli σ's, $\sigma_x = \begin{pmatrix} 0 & 1 \\ 1 & 0 \end{pmatrix}$ act on it which yields the vector $\begin{pmatrix} \cos(\theta/2) \\ \sin(\theta/2) \end{pmatrix}$. Hence if we rotate through $90^0$ we produce an eigen-vector of $\sigma_x$. Hence we can measure the x-component of the spin of the second particle this way. Now if we follow EPR we would say that we have made an implicit measurement of the z component, which in view of the way we did it, has "reality" and an explicit measurement of the x component. But $\sigma_x$ and $\sigma_z$ do not commute so we cannot have simultaneous knowledge of these two components. The solution here is as before. Implicit measurements do not count. If you try to make real measurements of these components the uncertainty principle limits what you can do. It is this version of the EPR that you will find in those text books that discuss the subject at all.

      John Bell was much taken both by Bohm's version of the EPR and by his pilot wave interpretation of the quantum theory. He put them together in a way in which we have looked at the foundations of the quantum theory every since. Let is imagine we have a source that produces a pair of spin-1/2 particles in a singlet state: (↑₁↓₂-↓₁↑₂)/√2. Let is also imagine that we have a pair of Stern-Gerlach magnets widely separated in space. Let us also suppose that we can rotate these magnets through various angles about the y-axis as we have discussed above. The spin-1/2 particles fly off towards their respective magnets. We have already discussed the anti-correlation of the spins when the magnets are pointing in the same direction. But suppose that is rotated through an a and the other through an angle b, what are the probabilities of finding the possible correlations? Some quantum mechanical maneuvering give the following answers:

P(up,up)=P(down.down)=1/2 sin((a-b)/2)$^2$

P(up,down)=P(down,up)=1/2-1/2 sin((a-b)/2)$^2$

$=1/2\cos((a-b)/2)^2$

We can now ask for the net correlation of the two spins which we define to be

P(up,up)+ P(down.down)- P(up,down)- P(down,up)

But this equals –cos(a-b). This is the quantum mechanical result. If Bohm's pilot wave interpretation reproduces this answer, how does it do it? Indeed, the pilot wave interpretation does reproduce this answer. In Bohmian mechanics there are two types of related equations. There is a first order set of equations that describe the motions of the particles. These motions are perfectly classical and deterministic. But they are guided by pilot waves that satisfy Schrödinger equations.  If there are two or more particles in interaction then these Schrödinger wave functions are no longer separable. What happens to the guide wave of one function depends on what happens to the other and vice versa. So if one particle's guide wave is influenced by the rotation of a Stern-Gerlach magnet this effects instantaneously the guide wave of the other. This is Einstein's "spooky actions at a distance" in its most blatant form. What Bell showed is that this kind of non-locality is an ineluctable part of the quantum theory. You can't derive the correlations without it. It is almost a century since Stern and Gerlach first showed experimentally how odd quantum mechanics is. Using their methods we now see that the theory is even odder than they imagined.